\documentclass[pra,aps,showpacs,twocolumn]{revtex4-1}
\usepackage{amsmath}
\usepackage{amssymb}
\usepackage{graphicx}
\usepackage{epstopdf}
\usepackage{easybmat}
\usepackage{CJK}

\usepackage{color,soul}
\usepackage{times,txfonts}
\setstcolor{blue}
\setulcolor{red}
\makeatletter

\newcommand{\Rmnum}[1]{\expandafter\@slowromancap\romannumeral #1@}
\makeatother
\begin{document}

\title{Nonadiabatic geometric quantum computation in decoherence-free subspaces based on unconventional geometric phases}
\author{P. Z. Zhao}
\affiliation{Department of Physics, Shandong University, Jinan 250100, China}
\author{G. F. Xu}
\email{Email: sduxgf@163.com}
\affiliation{Department of Physics, Shandong University, Jinan 250100, China}
\author{D. M. Tong}
\email{Email: tdm@sdu.edu.cn}
\affiliation{Department of Physics, Shandong University, Jinan 250100, China}
\date{\today}

\begin{abstract}
Nonadiabatic geometric quantum computation in decoherence-free subspaces has received increasing attention due to the merits of its high-speed implementation and robustness against both control errors and decoherence. However, all the previous schemes in this direction have been based on the conventional geometric phases, of which the dynamical phases need to be removed. In this paper, we put forward a scheme of nonadiabatic geometric quantum computation in decoherence-free subspaces based on unconventional geometric phases, of which the dynamical phases do not need to be removed.  Specifically, by using three physical qubits undergoing collective dephasing to encode one logical qubit, we realize a universal set of geometric gates nonadiabatically and unconventionally. Our scheme not only maintains all the merits of nonadiabatic geometric quantum computation in decoherence-free subspaces, but also avoids the additional operations required in the conventional schemes to cancel the dynamical phases.
\end{abstract}

\maketitle

\section{Introduction}

Quantum computation is based on quantum logic which is totally different from Boolean logic. This feature allows a quantum computer to solve many problems, such as factoring large integers \cite{Shor} and searching unsorted data \cite{Grover}, much more rapidly than a classical computer. To implement practical quantum computation, a universal set of quantum gates with extremely high fidelities needs to be realized. However, quantum errors inevitably affect quantum gates and make practical realizations difficult. To overcome this problem, geometric quantum computation (GQC) has been proposed. As we know, geometric gates depend only on evolutional paths and not on evolutional details so that it is robust against control errors, which are  regarded as one main obstacle in realizing quantum computation.

To realize geometric gates, various geometric phases have been used. Originally, adiabatic geometric phases \cite{Berry,Wilczek} were used to realize geometric gates and this kind of computation schemes was known as  adiabatic GQC \cite{Jones,Zanardi,Duan2001}. Although adiabatic GQC has geometric robustness, the evolution time associated with adiabatic requirement is usually longer than the coherence time and thus the practical computation is seriously collapsed. To overcome this problem, nonadiabatic GQC based on nonadiabatic and Abelian geometric phases \cite{Aharonov} was proposed \cite{Wang,Zhu2002}, and more interestingly nonadiabatic holonomic quantum computation, which is based on nonadiabatic and non-Abelian geometric phase \cite{Anandan}, was recently found \cite{E2012,Xu2012}. In realizing both adiabatic and nonadiabatic geometric gates, dynamical phases are usually removed and this kind of schemes is known as conventional GQC. However, the removal of dynamical phases is not necessary. Sometimes, total phases are proportional to geometric phases and the corresponding proportional coefficients are non-trivial constants or at least independent of some systematic parameters. In this case, total phases possess the same geometric robustness as geometric phases and can be directly used to realize geometric gates. By using this kind of total phases, unconventional GQC was realized \cite{Zhu2003}. Compared with conventional GQC, unconventional GQC eliminates some restrictions in practical realization, since the additional operations required to cancel the dynamical phases do not need to act on physical systems. Until now, both conventional GQC  and unconventional GQC have attracted much attention.

Although GQC is robust against control errors, decoherence as another main obstacle in realizing quantum computation is still detrimental to the computation. Thus, if one wants to further improve the fidelities of geometric gates, decoherence should be avoided. To do this, one promising way is to combine geometric gates with the strategies previously proposed to fight against decoherence. Considering that these strategies are not directly compatible with geometric gates, extra efforts are certainly needed to make the combination successful. Despite this, impressive progress has been made in this direction \cite{Wu2005,Zhang2006,Cen2006,Feng2009,Chen2012,Xu2012,Xu2014SR,Liang2014,Xue2015,Zhou,Xue2016,Ore2009PRL,
Oreshkov2009PRL,Ore2009PRA,Xu2014PRA,Zhang2014,Zheng2015,Sun2016} and many works have been done to realize GQC in decoherence-free subspaces (DFSs) \cite{Wu2005,Zhang2006,Cen2006,Feng2009,Chen2012,Xu2012,Xu2014SR,Liang2014,Xue2015,Zhou,Xue2016}.
Among these works, most of them realized conventional GQC in DFSs \cite{Wu2005,Zhang2006,Xu2012,Xu2014SR,Liang2014,Xue2015,Zhou,Xue2016}. Since unconventional GQC in DFSs shares all the robustness of conventional GQC in DFSs while avoiding the additional operations required to cancel the dynamical phases, realizing unconventional GQC in DFSs is of more practical importance. In fact, some works have made such attempts in the past few years \cite{Cen2006,Feng2009,Chen2012}. However, these works either adiabatically realized unconventional GQC in DFSs \cite{Cen2006} or only nonadiabatically realized a two-qubit unconventional geometric gate in DFSs \cite{Feng2009,Chen2012}. So far, unconventional nonadiabatic GQC in DFSs remains an open problem.

In this paper, we demonstrate how to nonadiabatically realize unconventional GQC in DFSs. Since we combine unconventional geometric phases and DFSs, obstacles of both control errors and decoherence are relaxed, and additional operations required to cancel the dynamical phases are not needed. Particularly, universality and nonadiabaticity, as two important improvements for realizing unconventional GQC in DFSs, are simultaneously achieved. In addition, the realization of our unconventional geometric gates does not utilize the usually used displacement operator, which may shed light on the applications of unconventional GQC.

\section{The Scheme}

Let us now elucidate our physical model. We consider a computational system consisting of $N$ qubits, with the Hamiltonian
\begin{align}
H_N=\sum_{k<l}\biggm(J^{xy}_{kl}R^{xy}_{kl}+J^{z}_{kl}R^{z}_{kl}\biggm)+\sum_{m}J^{z}_{m}\sigma^{z}_{m}, \label{hamiltonian}
\end{align}
where $J^{xy}_{kl}$ and $J^{z}_{kl}$ are the controllable coupling parameters, $J^{z}_{m}$ represents the effective local operation applying to the $m$th physical qubit, $\sigma^{\beta}_{\alpha}$ represents the Pauli $\beta$ operator acting on the $\alpha$th physical qubit, and $R^{xy}_{kl}$ and $R^{z}_{kl}$ are written as
\begin{align}
R^{xy}_{kl}=\frac{1}{2}\Bigr(\sigma^{x}_{k}\sigma^{x}_{l}+\sigma^{y}_{k}\sigma^{y}_{l}\Bigr), ~~~
R^{z}_{kl}=\sigma^{z}_{k}\sigma^{z}_{l}.
\end{align}
The Hamiltonian $H_N$ is the $XXZ$ Hamiltonian, which can be realized by a variety of physical systems. For example, in superconducting circuits, the first two terms can be realized by superconducting islands coupled to a ring by two symmetric Josephson junctions,  while the last term can be realized by a local operation acting on a single qubit \cite{Makhlin,Siewert,Averin}.

For the system considered here, the main source of decoherence is dephasing. If the $N$ qubits interact collectively with the environment, the interaction Hamiltonian can be written as
\begin{align}
H_{I}=\biggm(\sum_{k}\sigma^{z}_{k}\biggm)\otimes B, \label{interaction}
\end{align}
where $B$ is the operator on the common environment. For such a symmetric Hamiltonian $H_{I}$, DFSs can be found to fight against decoherence. In the following paragraphs, we will demonstrate how to realize a universal set of unconventional nonadiabatic geometric gates in these DFSs.

\subsection{One-logical-qubit geometric gates in DFSs}

To realize universal quantum computation, two noncommuting one-qubit gates and one nontrivial two-qubit gate are needed. First, we demonstrate how to realize the one-logical-qubit unconventional nonadiabatic geometric gates in DFSs.

Consider three physical qubits interacting collectively with the dephasing environment. For convenience, we denote the three physical qubits as 1, 2, and 3,respectively. With the interaction Hamiltonian in Eq. (\ref{interaction}), there exists a three-dimensional DFS
\begin{align}
\mathcal {S}={\it Span}\{|100\rangle,|010\rangle,|001\rangle\}.
\end{align}
Here, we use three-physical-qubit states in $\mathcal {S}$ to encode one-logical-qubit states and the specific encoding is $|0\rangle_{L}=|010\rangle$,  $|1\rangle_{L}=|001\rangle$, and $|a\rangle_L=|100\rangle$, where $|0\rangle_{L}$ and $|1\rangle_{L}$ are the computational basis of the logical qubit and $|a\rangle_L$ is used as an ancillary logical state.

To start with, we demonstrate how to realize the first one-logical-qubit unconventional nonadiabatic geometric gate
\begin{align}
U^{Z}_{L}=e^{i\gamma_{1}Z_{L}/2},
\end{align}
where $Z_L=|0\rangle_{LL}\langle0|-|1\rangle_{LL}\langle1|$ can be seen as the logical Pauli $Z$ operator and $\gamma_{1}$ is the corresponding rotation angle.
To realize the gate $U^{Z}_{L}$, we consider the Hamiltonian $H_{N=3}$. Specifically, we set
\begin{align}
J^{xy}_{13}=J_{1}\cos\theta,~~~J^{z}_{3}=-J_{1}\sin\theta,
\end{align}
and other coupling parameters to zero, where $J_{1}$ can be seen as the envelope of parameters $J^{xy}_{13}$ and $J^{z}_{3}$, and $\theta$ determines their relative strengths. After choosing the parameters as above, the Hamiltonian $H_{N=3}$ can be rewritten as
\begin{align}
\mathcal {H}_{1}=&J_{1}\cos\theta\Big(|a\rangle_{LL}\langle1|+|1\rangle_{LL}\langle a|\Big)+2J_{1}\sin\theta|1\rangle_{LL}\langle1|. \label{hamiltonian1}
\end{align}
Here, we have used the fact that the operator $|a\rangle_{LL}\langle a|+|0\rangle_{LL}\langle 0|+|1\rangle_{LL}\langle 1|$ is an identity operator and can be ignored because it only generates a global phase during evolution. With the expression of Hamiltonian $\mathcal {H}_{1}$, we can work out the corresponding evolution operator. By choosing the evolution period such that
\begin{align}
J_{1}\tau_{1}=2\pi,
\end{align}
the corresponding evolution operator in the basis $\{|a\rangle_L,|1\rangle_{L},|0\rangle_{L}\}$ reads
\begin{align}
U_{1}(\tau_1)=\left(
  \begin{array}{ccc}
   e^{-i\gamma_{1}} & 0 & 0\\
   0 & e^{-i\gamma_{1}} & 0\\
   0 & 0 & 1\\
  \end{array}
\right),
\end{align}
where the phase $\gamma_{1}$ is
\begin{align}
\gamma_{1}=2\pi\sin\theta. \label{global1}
\end{align}
Accordingly, the logical gate acting on the computational subspace spanned by $\{|0\rangle_{L},|1\rangle_{L}\}$ reads
\begin{align}
U_{1}^{\prime}(\tau_1)=|0\rangle_{LL}\langle0|+e^{-i\gamma_{1}}|1\rangle_{LL}\langle1|,
\end{align}
and the action of $U_{1}^{\prime}(\tau_1)$ is equivalent to that of $U^{Z}_{L}$. Furthermore, the evolution operator $U_{1}^{\prime}(\tau_1)$ is protected by DFS $\mathcal {S}$ all the time. Thus, we have nonadiabatically realized the one-logical-qubit gate $U^{Z}_{L}$ in DFS $\mathcal {S}$.

To demonstrate the geometric robustness of the logical gate $U^{Z}_{L}$, one usually needs to verify that the phases accumulated by states $|0\rangle_L$ and $|1\rangle_L$ have geometric robustness. However, the logical state $|0\rangle_L$ is decoupled from evolution all the time. So, the above verification is reduced and only the phase accumulated by state $|1\rangle_L$ needs to be examined. To this end, we calculate the corresponding dynamical and geometric phases, and then investigate their features. The dynamical phase accumulated by state $|1\rangle_L$ reads
\begin{align}
\gamma_{d_{1}}&=-\int^{\tau_1}_{0}\langle 1|_LU_1^{\dag}(t)\mathcal {H}_{1}U_1(t)|1\rangle_L dt\notag\\
&=-4\pi\sin\theta, \label{dp1}
\end{align}
where $U_1(t)=\exp({-i\mathcal {H}_{1}t})$. After getting the total phase $\gamma_1$ and dynamical phase $\gamma_{d_{1}}$, the corresponding geometric phase can be directly written as
\begin{align}
\gamma_{g_{1}}&=\gamma_{1}-\gamma_{d_{1}} \notag\\
&=6\pi\sin\theta. \label{gp1}
\end{align}
According to Eqs. (\ref{dp1}) and (\ref{gp1}), the dynamical phase is proportional to the corresponding geometric phase and the proportional coefficient reads
\begin{align}
\frac{\gamma_{d_{1}}}{\gamma_{g_{1}}}=-\frac{2}{3}.
\end{align}
Since the proportional coefficient is constant, the realized gate $U_{L}^Z$ is a one-logical-qubit unconventional nonadiabatic geometric gate in DFS $\mathcal {S}$.

Now, we demonstrate how to realize the second one-logical-qubit gate
\begin{align}
U^{X}_{L}=e^{i\gamma_{2}X_L/2},
\end{align}
where $X_L=|0\rangle_{LL}\langle1|+|1\rangle_{LL}\langle0|$ can be seen as the logical Pauli $X$ operator, and $\gamma_{2}$ is the rotation angle. To this end, we set the non-zero parameters of Hamiltonian $H_{N=3}$ described by Eq. (\ref{hamiltonian}) as
\begin{align}
&\sqrt{2}J^{xy}_{12}=-\sqrt{2}J^{xy}_{13}=J_{2}\cos\varphi,\notag\\
&J^{xy}_{23}=2J^{z}_{2}=2J^{z}_{3}=-J_{2}\sin\varphi,
\end{align}
where $J_{2}$ can be seen as the envelope of parameters $J^{xy}_{12}$, $J^{xy}_{13}$, $J^{xy}_{23}$, $J^{z}_{2}$, and $J^{z}_{3}$, and $\varphi$ determines their relative strengths. Then, the Hamiltonian $H_{N=3}$ described by Eq. (\ref{hamiltonian}) can be rewritten as
\begin{align}
\mathcal {H}_{2}=J_{2}\cos\varphi\Big(|a\rangle_{LL}\langle-|+|-\rangle_{LL}\langle a|\Big)+2J_{2}\sin\varphi|-\rangle_{LL}\langle -|. \label{hamiltonian2}
\end{align}
The logical state $|-\rangle_{L}$ and its orthogonal logical state $|+\rangle_{L}$ are written as
\begin{align}
|-\rangle_{L}=&\frac{1}{\sqrt{2}}(|0\rangle_{L}-|1\rangle_{L}), \notag\\
|+\rangle_{L}=&\frac{1}{\sqrt{2}}(|0\rangle_{L}+|1\rangle_{L}).
\end{align}
By choosing the evolution period of the quantum system such that
\begin{align}
J_{2}\tau_{2}=2\pi,
\end{align}
the corresponding evolution operator in the basis $\{|a\rangle_L,|-\rangle_{L},|+\rangle_{L}\}$ reads
\begin{align}
U_{2}(\tau_2)=\left(
  \begin{array}{ccc}
   e^{-i\gamma_{2}} & 0 & 0\\
   0 & e^{-i\gamma_{2}} & 0\\
   0 & 0 & 1\\
  \end{array}
\right),
\end{align}
where the phase $\gamma_{2}$ is
\begin{align}
\gamma_{2}=2\pi\sin\varphi. \label{global2}
\end{align}
Clearly, the logical gate acting on the computational subspace spanned by $\{|0\rangle_{L},|1\rangle_{L}\}$ reads
\begin{align}
U_{2}^\prime(\tau_2)=|+\rangle_{LL}\langle+|+e^{-i\gamma_{2}}|-\rangle_{LL}\langle-|,
\end{align}
and the action of $U_{2}^\prime(\tau_2)$ is equivalent to that of $U_L^X$. So, we have nonadiabatically realized one-logical-qubit gates $U^{X}_{L}$ in the subspace $\mathcal {S}$, and in the following, we will demonstrate its geometric robustness.

Similarly to the case of the first one-logical-qubit gate, we need to verify only that the phase accumulated by the state $|-\rangle_L$ has geometric robustness. As one can see, the dynamical phase accumulated by the state $|-\rangle_L$ reads
\begin{align}
\gamma_{d_{2}}&=-\int^{\tau_2}_{0}\langle -|_LU^\dag_2(t)\mathcal {H}_{2}U_2(t)|-\rangle_L dt\notag\\
&=-4\pi\sin\varphi, \label{dp2}
\end{align}
where $U_2(t)=\exp({-i\mathcal {H}_{2}t})$. Then, the corresponding geometric phase accumulated by the state $|-\rangle_L$ reads
\begin{align}
\gamma_{g_{2}}&=\gamma_{2}-\gamma_{d_{2}} \notag\\
&=6\pi\sin\varphi. \label{gp2}
\end{align}
Clearly, no matter what values the parameters take, the dynamical phase is proportional to the geometric phase and the proportional coefficient is constant,
\begin{align}
\frac{\gamma_{d_{2}}}{\gamma_{g_{2}}}=-\frac{2}{3}.
\end{align}
So, the realized gate $U_{L}^X$ is a one-logical-qubit unconventional nonadiabatic geometric gate in DFS $\mathcal {S}$. Furthermore, it is easy to verify that gates $U_{L}^Z$ and $U_{L}^X$ are two noncommuting gates, by which arbitrary one-qubit gates can be realized.

\subsection{Two-logical-qubit geometric gate in DFSs}

In the above section, we have realized two noncommuting one-logical-qubit gates in DFSs. To realize universal quantum computation, we still need to realize a two-logical-qubit unconventional nonadiabatic geometric gate in DFSs. In this section, we demonstrate how to do this.

Consider two logical qubits, each of which contains three physical qubits. For convenience, we denote the six physical qubits 1, 2, 3, 4, 5 and 6,respectively. Suppose these six physical qubits interact collectively with a dephasing environment. Then there exists a six-dimensional DFS
\begin{align}
{\mathcal {S}}^\prime=&{\it Span}\{|010010\rangle,|010001\rangle,|001010\rangle,\notag\\
&|001001\rangle,|011000\rangle,|000011\rangle\}.
\end{align}
To be compatible with the one-logical-qubit encoding, we encode the two-qubit logical states as $|00\rangle_{L}=|010010\rangle$, $|01\rangle_{L}=|010001\rangle$, $|10\rangle_{L}=|001010\rangle$, and $|11\rangle_{L}=|001001\rangle$. Meanwhile, we use the two states $|a_{1}\rangle_L=|011000\rangle$ and $|a_{2}\rangle_L=|000011\rangle$ as two ancillary states. To realize the two-logical-qubit gate, we consider the Hamiltonian $H_{N=6}$ and set the corresponding nonzero parameters as
\begin{align}
J^{xy}_{35}=&{J}_3\cos\phi,~~~J^{z}_{36}={J}_3\sin\phi,
\end{align}
where ${J}_3$ can be seen as the envelop of parameters $J^{xy}_{35}$ and $J^{z}_{36}$, and $\phi$ determines the relative strengths of these parameters. Then the Hamiltonian $H_{N=6}$ in Eq. (\ref{hamiltonian}) can be written as
\begin{align}
{\mathcal {H}}_3=&{J}_3\cos\phi\biggm(|a_{1}\rangle_{LL}\langle 00|+|a_{2}\rangle_{LL}\langle 11|+H.c.\biggm)\notag\\ &+2{J}_3\sin\phi\biggm(|00\rangle_{LL}\langle 00|+|11\rangle_{LL}\langle 11|\biggm). \label{hamiltonian3}
\end{align}
If the evolution period is chosen to satisfy
\begin{align}
J_3{\tau}_3=2\pi,
\end{align}
the corresponding evolution operator in the basis $\{|a_{1}\rangle_L,|a_{2}\rangle_L,|00\rangle_{L},|01\rangle_{L},|10\rangle_{L},|11\rangle_{L}\}$ reads
\begin{align}
{U}_3({\tau}_3)=\left(
  \begin{array}{cccccc}
   e^{-i\gamma_3} & 0 & 0 & 0 & 0 & 0 \\
   0 & e^{-i\gamma_3} & 0 & 0 & 0 &0 \\
   0 & 0 & e^{-i\gamma_3} & 0 & 0 & 0 \\
   0 & 0 & 0 & 1 & 0 & 0 \\
   0 & 0 & 0 & 0 & 1 & 0 \\
   0 & 0 & 0 &0 & 0 & e^{-i\gamma_3} \\
  \end{array}
\right),
\end{align}
where the phase $\gamma_3$ is
\begin{align}
{\gamma}_3=2\pi\sin\phi.
\end{align}
According to the above equations, the computational subspace evolves cyclically and the corresponding logical operator reads
\begin{align}
{U}^\prime_{3}({\tau}_3)=&|01\rangle_{LL}\langle01|+|10\rangle_{LL}\langle10|\notag\\
&+e^{-i{\gamma}_3}\Bigm(|00\rangle_{LL}\langle00|+|11\rangle_{LL}\langle11|\Bigm),
\end{align}
One can verify that the logical gate ${U}^\prime_{3}({\tau}_3)$ is an entangling gate if $\gamma_3\neq0,\pm\pi$. Therefore, we have nonadiabatically realized an entangling two-logical-qubit gate in DFS $\mathcal {S}^\prime$.

By observing the Hamiltonian $\mathcal {H}_3$ in Eq. (\ref{hamiltonian3}), one can see that the logical states $|01\rangle_{L}$ and $|10\rangle_{L}$ are decoupled from the evolution. Thus, to ensure the geometric robustness  of gate ${U}^\prime_{3}({\tau}_3)$, only the phases accumulated by states $|00\rangle_{L}$ and $|11\rangle_{L}$ need to be discussed. Through direct calculation, the states $|00\rangle_{L}$ and $|11\rangle_{L}$ acquire the same dynamical phases which can be expressed as
\begin{align}
\gamma_{d_3}&=-\int^{\tau_3}_{0}\langle 00|_{L}U_3^\dag(t){\mathcal {H}}_3U_3(t)|00 \rangle_{L} dt\notag\\
&=-\int^{\tau_3}_{0}\langle 11|_{L}U_3^\dag(t){\mathcal {H}}_3U_3(t)|11\rangle_{L} dt\notag\\
&=-4\pi\sin\phi.
\end{align}
Here, $U_3(t)=\exp({-i\mathcal {H}_3 t})$. Considering that the total phases accumulated by states $|00\rangle_{L}$ and $|11\rangle_{L}$ are also the same, the geometric phases accumulated by these two states have the same value and can be expressed as
\begin{align}
\gamma_{g_3}&=\gamma_3-\gamma_{d_3}\notag\\
&=6\pi\sin\phi.
\end{align}
According to the above discussion, the dynamical phases are proportional to the corresponding geometric phases with the same proportional coefficients
\begin{align}
\frac{\gamma_{d_3}}{\gamma_{g_3}}=-\frac{2}{3}.
\end{align}
Again, the above coefficients are constants and then the realized gate $U_3^\prime(\tau_3)$ is an unconventional nonadiabatic geometric gate protected by DFS $\mathcal {S}^\prime$ all the time.

\section{Conclusion}

In conclusion, we have proposed a scheme of nonadiabatic GQC in DFSs based on unconventional geometric phases. Specifically, we use three physical qubits undergoing collective dephasing to encode one logical qubit, and further realize a universal set of geometric gates in DFSs nonadiabatically and unconventionally. Similarly to the schemes of nonadiabatic holonomic quantum computation in DFSs or noiseless subsystems \cite{Xu2012,Liang2014,Xue2015,Zhou,Xue2016,Zhang2014}, our scheme uses Hamiltonians with three-level structures. However, the dynamical phases of our scheme are proportional to the total phases, while
the dynamical phases of the schemes of nonadiabatic holonomic quantum computation in DFSs or noiseless subsystems are equal to 0.
Our quantum computation scheme combines the advantages of the previous schemes of nonadiabatic GQC in DFSs, such as the high-speed implementation and the robustness against control errors and decoherence, and the advantage of unconventional geometric gates not needing to cancel the dynamical phases. Our scheme is based on the $XXZ$ Hamiltonian and can be realized in various realistic quantum systems. Besides, the realization of our unconventional geometric gates does not utilize the usually used displacement operator. We hope that our scheme will shed light on the applications of unconventional GQC in DFSs.

\begin{acknowledgments}
PZZ acknowledges the support of National Natural Science Foundation of China through Grant No. 11575101. GFX acknowledges the support of National Natural Science Foundation of China through Grant No. 11547245 and No. 11605104, and of the Future Project for Young Scholars of Shandong University through Grant No. 2016WLJH21. DMT acknowledges the support of the National Basic Research Program of China through Grant No. 2015CB921004.
\end{acknowledgments}

\end{document}